\documentclass[12pt, onecolumn, draftclsnofoot, journal]{IEEEtran}

\usepackage{amsmath,amsfonts}
\usepackage{algorithmic}
\usepackage{algorithm}
\usepackage{array}
\usepackage[caption=false,font=small]{subfig}
\usepackage{textcomp}
\usepackage{stfloats}
\usepackage{url}
\usepackage{verbatim}
\usepackage{graphicx}
\usepackage[dvipsnames]{xcolor}
\usepackage{enumitem}
\usepackage{rotating}
\usepackage{comment}
\usepackage{pdflscape}
\usepackage{setspace}
\usepackage{footmisc}
\usepackage{cite}
\usepackage{tikz}
\definecolor{matlabblue}{RGB}{0,0,255}
\usepackage[acronym,shortcuts]{glossaries}

\newacronym{1G}{1G}{first generation}
\newacronym{2G}{2G}{second generation}
\newacronym{3G}{3G}{third generation}
\newacronym{4G}{4G}{fourth generation}
\newacronym{5G}{5G}{fifth generation}
\newacronym{6G}{6G}{sixth generation}
\newacronym{6G+}{6G+}{6G and beyond}
\newacronym{CDMA}{CDMA}{code-division multiple access}
\newacronym{OFDM}{OFDM}{orthogonal frequency-division multiplexing}
\newacronym{OFDMA}{OFDMA}{orthogonal frequency-division multiple access}
\newacronym{LEO}{LEO}{low Earth orbit}
\newacronym{SAGIN}{SAGIN}{space-air-ground integrated network}
\newacronym{ISAC}{ISAC}{integrated sensing and communications}
\newacronym{LTI}{LTI}{linear time-invariant}
\newacronym{LTV}{LTV}{linear time-varying}
\newacronym{IoT}{IoT}{Internet of Things}
\newacronym{AFDM}{AFDM}{affine frequency-division multiplexing}
\newacronym{AFDMA}{AFDMA}{affine frequency-division multiple access}
\newacronym{ChDMA}{ChDMA}{chirp-division multiple access}
\newacronym{LoRa}{LoRa}{long range}
\newacronym{radar}{radar}{radio detection and ranging}
\newacronym{V2X}{V2X}{vehicle-to-everything}
\newacronym{SNR}{SNR}{signal-to-noise ratio}
\newacronym{EHF}{EHF}{extremely high frequency}
\newacronym{MIMO}{MIMO}{multiple-input multiple-output}
\newacronym{FFT}{FFT}{fast Fourier transform}
\newacronym{IFFT}{IFFT}{inverse fast Fourier transform}
\newacronym{DFT}{DFT}{discrete Fourier transform}
\newacronym{NB-IoT}{NB-IoT}{narrowband IoT}
\newacronym{DAFT}{DAFT}{discrete affine Fourier transform}
\newacronym{IDAFT}{IDAFT}{inverse discrete affine Fourier transform}
\newacronym{UAV}{UAV}{unmanned aerial vehicle}
\newacronym{mmWave}{mmWave}{millimeter-wave}
\newacronym{THz}{THz}{Terahertz}
\newacronym{OTFS}{OTFS}{orthogonal time frequency space}
\newacronym{PSD}{PSD}{power spectral density}
\newacronym{ODDM}{ODDM}{orthogonal delay-Doppler division multiplexing}
\newacronym{OCDM}{OCDM}{orthogonal chirp-division multiplexing}
\newacronym{SFFT}{SFFT}{symplectic finite Fourier transform}
\newacronym{ISFFT}{ISFFT}{inverse symplectic finite Fourier transform}
\newacronym{IDZT}{IDZT}{inverse discrete Zak transform}
\newacronym{DD}{DD}{delay-Doppler}
\newacronym{TF}{TF}{time-frequency}
\newacronym{FMCW}{FMCW}{frequency-modulated continuous wave}
\newacronym{ICI}{ICI}{inter-carrier interference}
\newacronym{ISI}{ISI}{inter-symbol interference}
\newacronym{CP}{CP}{cyclic prefix}
\newacronym{CPP}{CPP}{chirp periodic prefix}
\newacronym{TDMA}{TDMA}{time-division multiple access}
\newacronym{3GPP}{3GPP}{3rd Generation Partnership Project}
\newacronym{NR}{NR}{New Radio}
\newacronym{MMSE}{MMSE}{minimum mean square error}
\newacronym{MP}{MP}{message passing}
\newacronym{BER}{BER}{bit error rate}
\newacronym{FDMA}{FDMA}{frequency-division multiple access}
\newacronym{OOBE}{OOBE}{out-of-band emissions}
\newacronym{RB}{RB}{resource block}
\newacronym{BWP}{BWP}{bandwidth part}
\newacronym{URLLC}{URLLC}{ultra-reliable low-latency communication}
\newacronym{FPGA}{FPGA}{field-programmable gate array}
\newacronym{ASIC}{ASIC}{application-specific integrated circuit}
\newacronym{AF}{AF}{ambiguity function}
\newacronym{ISLR}{ISLR}{integrated sidelobe ratio}
\newacronym{PSLR}{PSLR}{peak-to-sidelobe ratio}
\newacronym{CSI}{CSI}{channel state information}
\newacronym{DFnT}{DFnT}{discrete Fresnel transform}
\newacronym{RRC}{RRC}{root raised cosine}
\newacronym{PN}{PN}{pseudo-noise}
\newacronym{DFT-s-OFDM}{DFT-s-OFDM}{discrete Fourier transform-spread-OFDM}
\newacronym{PAPR}{PAPR}{peak-to-average power ratio}
\newacronym{mMIMO}{mMIMO}{massive MIMO}
\newacronym{mMTC}{mMTC}{massive machine-type communication}
\newacronym{eMBB}{eMBB}{enhanced mobile broadband}
\newacronym{AI}{AI}{artificial intelligence}
\newacronym{UWB}{UWB}{ultra-wideband}
\newacronym{KPI}{KPI}{key performance indicator}
\newacronym{IIoT}{IIoT}{industrial Internet of Things}
\newacronym{QAM}{QAM}{quadrature amplitude modulation}
\newacronym{NEW}{NEW}{next-evolution waveforms}
\newacronym{NTN}{NTN}{non-terrestrial network}
\newacronym{RIS}{RIS}{reconfigurable intelligent surface}
\newacronym{XL-MIMO}{XL-MIMO}{extremely large-scale MIMO}
\newacronym{FAS}{FAS}{fluid antenna system}
\newacronym{RAQR}{RAQR}{Rydberg atomic quantum receiver}
\newacronym{SCMA}{SCMA}{sparse code multiple access}
\newacronym{IM}{IM}{index modulation}
\newacronym{PLS}{PLS}{physical-layer security}
\newacronym{DDDMA}{DDDMA}{delay-Doppler domain multiple access}
\newacronym{IQ}{IQ}{in-phase and quadrature}
\newacronym{LMMSE}{LMMSE}{linear minimum mean square error}
\newacronym{CFO}{CFO}{carrier frequency offset}
\newacronym{FBMC}{FBMC}{filter bank multicarrier}
\newacronym{CRLB}{CRLB}{Cram\'{e}r-Rao lower bound}
\newacronym{RAN}{RAN}{radio access network}
\newacronym{SLM}{SLM}{selected mapping}
\newacronym{CDD}{CDD}{cyclic delay diversity}
\newacronym{LTE}{LTE}{Long-Term Evolution}
\newacronym{SIM}{SIM}{stacked intelligent surface}
\newacronym{REMS}{REMS}{reconfigurable electromagnetic structure}
\newacronym{RF}{RF}{radio frequency}
\newacronym{MC}{MC}{multicarrier}
\newcommand{\rev}[1]{#1}%
\definecolor{revredink}{rgb}{0.80,0,0}

\definecolor{revblue}{rgb}{0,0,0.85}

\makeatletter
\@for\@revbibkey:={}\do{%
  \expandafter\global\expandafter\let
    \csname revbib@\@revbibkey\endcsname\@empty}
\let\revbib@bibitem\@bibitem
\def\@bibitem#1{%
  \color{black}\revbib@bibitem{#1}}
\makeatother

\begin{document}

\title{The Resurrection of Spectrum Spreading for 6G and Beyond: From Sinusoids to Chirps}

\author{
Hyeon Seok Rou,~\IEEEmembership{Member,~IEEE}, \\
Giuseppe Thadeu Freitas de Abreu,~\IEEEmembership{Senior Member,~IEEE}, \\
Emil Bj\"{o}rnson,~\IEEEmembership{Fellow,~IEEE}, \\
Sunwoo Kim,~\IEEEmembership{Senior Member,~IEEE}, \\
and Marios Kountouris,~\IEEEmembership{Fellow,~IEEE}.
}

\maketitle

\begin{abstract}

\Ac{OFDM} and its \rev{sinusoidal} subcarriers have underpinned the \acs{4G} and \acs{5G} eras, delivering high spectral efficiency and resilience to multipath fading through an efficient multicarrier architecture.
However, as future systems move toward doubly dispersive environments driven by high-mobility applications and migration to \acs{mmWave}/sub-\acs{THz} bands, the time-invariance assumption underlying \ac{OFDM} becomes increasingly difficult to maintain, and Doppler-induced degradation becomes prominent.
While enhancements such as \ac{MIMO}, advanced coding, and scheduling provide incremental remedies, they introduce additional overhead, because the sinusoidal subcarrier itself offers no inherent waveform-level robustness to Doppler impairments. 
Accordingly, two time-frequency spreading philosophies have emerged to improve Doppler resilience by distributing each symbol's energy across both dimensions of the time-frequency plane: (i) 2D isotropic spreading via the \ac{DD} domain, exemplified by the \ac{OTFS} family, and (ii) sheared spreading via parametrizable chirps, exemplified by the \ac{AFDM} family. 
In this article, we examine key considerations for future waveform design across these paradigms and argue that transitioning from the sinusoidal subcarriers of \ac{OFDM} to the chirp-based subcarriers offers a viable design direction for improving Doppler robustness while retaining much of the mature \ac{OFDM} infrastructure. 
This perspective also highlights the suitability of chirp-based waveforms for \ac{ISAC} and their extensibility to emerging physical-layer techniques.
Overall, we argue that the transition \textit{from sinusoids to chirps} is a technically motivated, compelling
evolutionary direction for future wireless physical layer design.
\end{abstract}

\glsresetall

\newpage

$~$ \vspace{-1ex}
\section*{The Rate-Robustness Pendulum of Wireless Evolution}
\label{sec:introduction}
$~$ \vspace{-2ex}

The evolution of wireless cellular systems is often oversimplified as a monotonic trajectory toward ever-higher data rates, with the gradual inclusion of applications and features, as illustrated in Fig.~\ref{fig:trend_linear}: from analog voice connectivity in the \ac{1G}, to digital traffic in the \ac{2G}, mobile internet in the \ac{3G}, broadband connectivity in the \ac{4G}, and the flexibility of \ac{5G}, which supports \ac{mMTC}, \ac{URLLC}, and \ac{eMBB} \cite{ITU_M2083}. More recently, \ac{6G} systems have been envisioned around \acs{AI}-native operation and \ac{ISAC}~\cite{Liu_VTM25}.

A closer examination\rev{, however,} reveals that this evolution is more accurately represented \rev{as a} three-dimensional trajectory in the joint \textit{rate-robustness-time} space, as depicted in Fig.~\ref{fig:trend_3d}.
The conventional rate-centric narrative is \rev{just a projection of the true richer trend in 3D, which} obscures the oscillating design priority along the robustness axis.
\rev{This new robustness axis exposes a} pendulum-like engineering rhythm between the two design objectives\rev{:} \ac{2G} pursued robustness against transmission errors through digitalization and coding, \ac{3G} pursued robustness against interference through wideband spectrum spreading, \ac{4G} prioritized mobile broadband rates through narrowband multiplexing, and \ac{5G} further enhanced rate and flexibility through \ac{MIMO} and advanced signal processing.

As we approach \ac{6G+}, however, \rev{a revised priority on robustness is required, driven by the demands of emerging} \ac{6G} applications such as \ac{UAV} and \ac{LEO} satellite networks, \ac{IIoT}, and high-speed \ac{V2X} \cite{Liu_VTM25}.
To motivate this architectural realignment, we cast the wireless evolution of the past decades as a \textit{dialectic rhythm} composed of the following stages: \vspace{0.5ex}
\begin{itemize}
     \item \textbf{\textit{Thesis}}: The era of interference robustness through spectrum spreading.
     \item \textbf{\textit{Antithesis}}: The era of rate and efficiency through multiplexing and \ac{MIMO}.
     \item \textbf{\textit{Crisis}}: The Doppler bottleneck arising from high-mobility and high-frequency bands.
     \item \textbf{\textit{Synthesis}}: The era of Doppler robustness through the reintroduction of spectrum spreading.
 \end{itemize}

\vspace{-3ex}

\clearpage
\begin{landscape}
\begin{center}
\end{center}
\vspace{-1ex}
\begin{figure}[H]
\centering
\subfloat[Conventional rate-centric representation.]{\includegraphics[width=0.445\columnwidth]{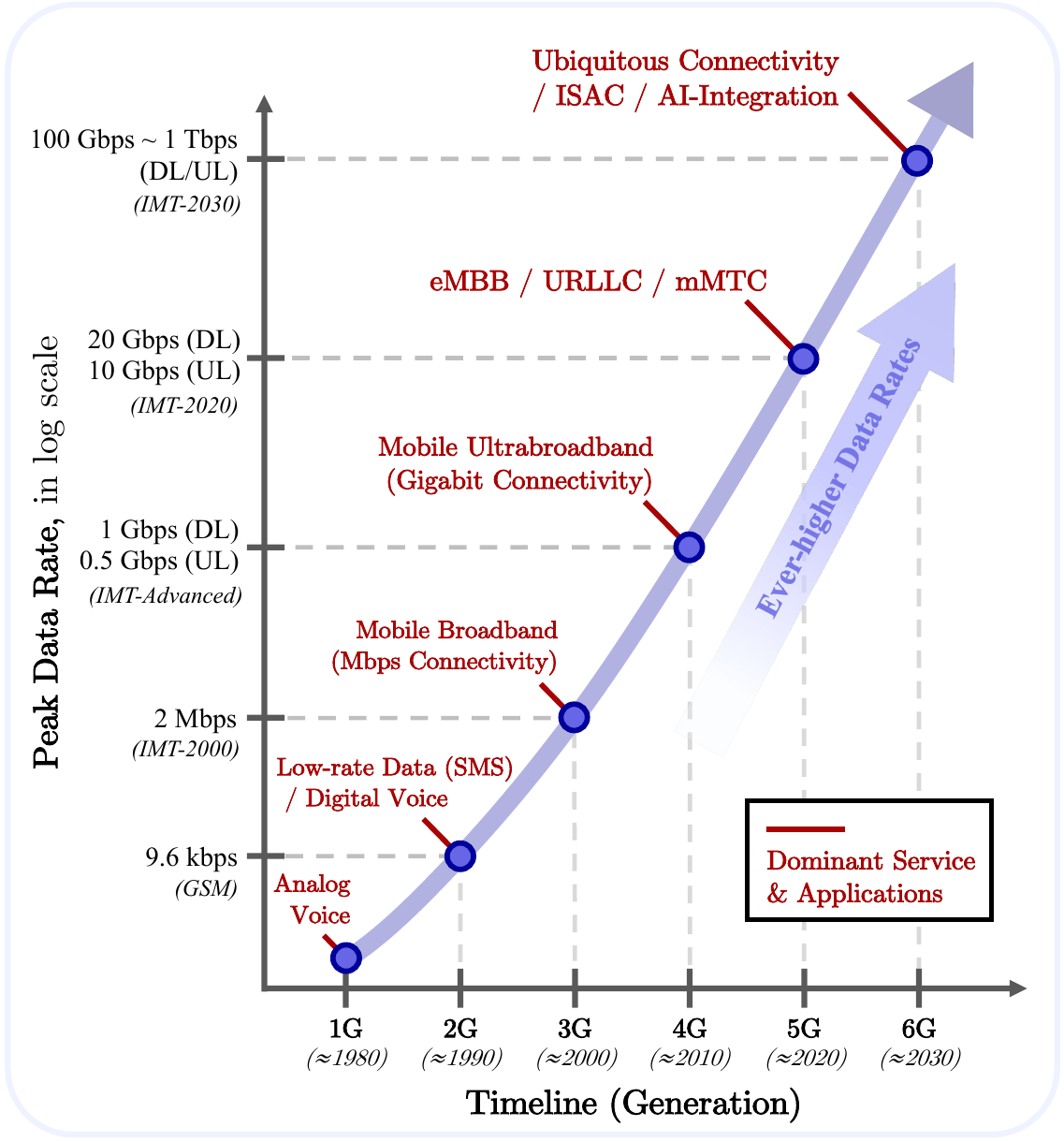}
\label{fig:trend_linear}}
\hfill
\subfloat[Extended three-dimensional representation in rate, robustness, and time.]{\includegraphics[width=0.54\columnwidth]{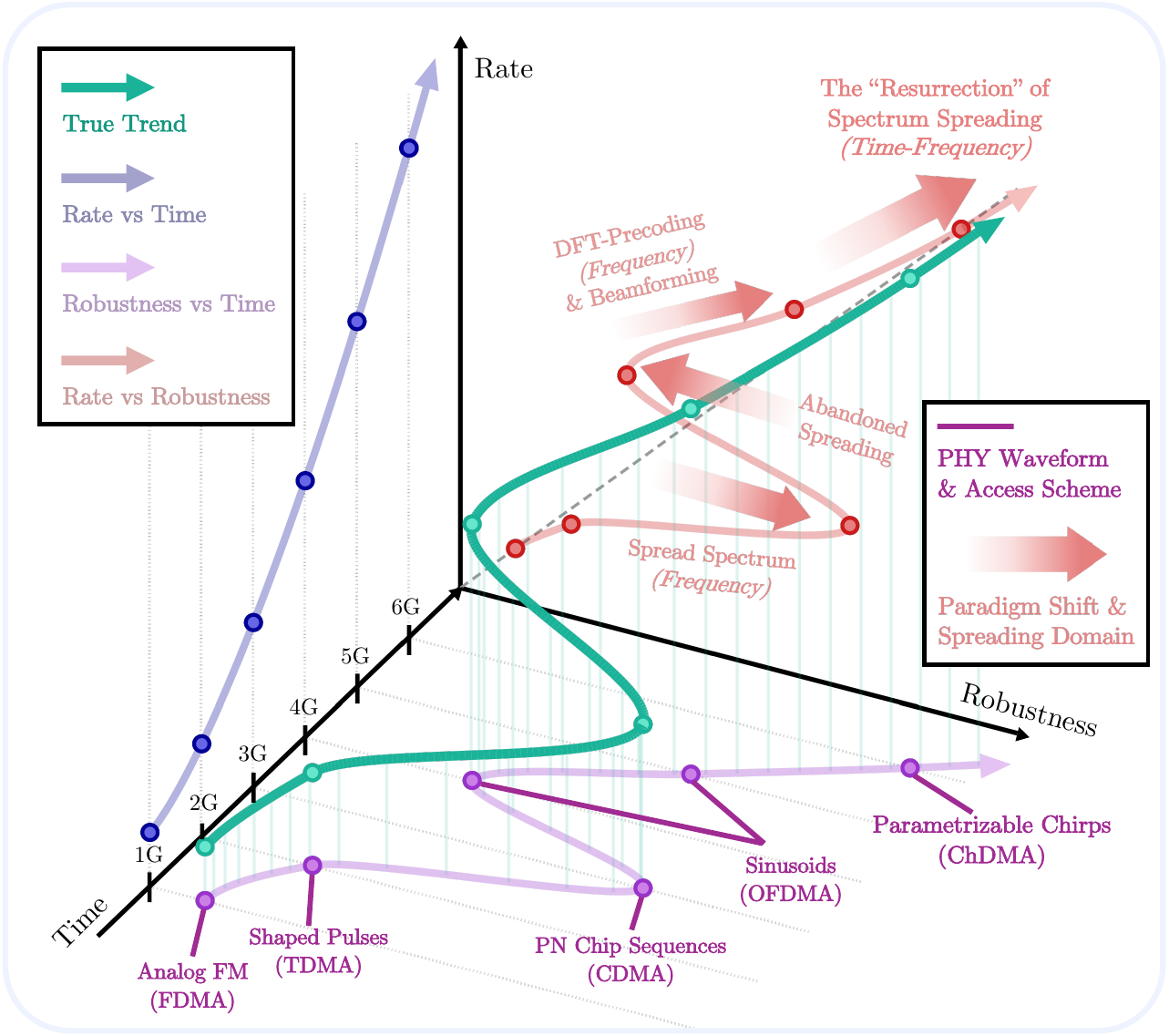}%
\label{fig:trend_3d}}
\caption{Two complementary representations of the wireless physical layer evolution leading up to \acs{6G} \rev{(conceptual illustration, axes not to scale)}.
(a)~the conventional single objective view emphasizing exponentially increasing peak data rates, and 
(b)~the extended perspective which reveals the pendulum-like oscillating priorities between the two key engineering philosophies of robustness and rate, against time, and it can be seen that the rate-centric view is only a projection of the interpreted trend in the three dimensional rate-robustness-time paradigm.
\rev{The robustness axis denotes the dominant impairment targeted by each generation rather than a single metric: transmission-error robustness via digitalization and coding in \acs{2G}, interference robustness via spectrum spreading in \acs{3G}, multipath-fading robustness via multicarrier equalization in \acs{4G}/\acs{5G} (though secondary to rate), and Doppler/double-dispersion robustness via joint time-frequency spreading toward \acs{6G}.}}
\label{fig:trend_twoviews}
\end{figure}
\end{landscape}
\clearpage

\section*{The Dialectic Rhythm of Wireless Evolution}
\label{sec:dialectic_rhythm}

\subsection*{\textbf{The Thesis (\acs{3G}) -- The Era of Interference Robustness}}

As the design objective shifted toward universal connectivity in the early 2000s, \ac{3G} addressed the challenge of multiuser interference by anchoring its architecture in spread-spectrum principles.
In this framework, each user's narrowband information signal was multiplied in time by a \ac{PN} chip sequence, flattening and widening the signal's \ac{PSD} across the available bandwidth, implementing \textit{spectrum spreading} \cite{Viterbi_CDMA1995}.
This spreading served two purposes.
First, the \ac{CDMA} scheme was introduced, whereby assigning each user a unique semi-orthogonal \ac{PN} sequence enabled simultaneous transmission by multiple users over the same bandwidth.
In turn, matched-filter despreading at the receiver recovers the desired user's signal efficiently while enabling multi-user interference suppression via the so-called processing gain.
From a waveform-centric perspective, \rev{this wideband occupancy implies} {inherent frequency diversity}: the signal averages over fading realizations, providing intrinsic resilience to frequency-selective fading without external mechanisms.

\subsection*{\textbf{The Antithesis (\acs{4G}/\acs{5G}) -- The Era of Rate and Efficiency}}

With the advent of smartphones and mobile Internet services, the design mandate shifted toward achieving higher data rates, for broadband access in \ac{4G} and for heterogeneous services, including \ac{eMBB}, \ac{URLLC}, and \ac{mMTC}, in \ac{5G} \cite{ITU_M2083}.
This motivated the search for alternatives to \ac{CDMA}, whose capacity was fundamentally limited by its spreading-code design and therefore insufficient for the demands of the new era.

As a solution, \ac{OFDM} \cite{Nee_OFDM2000} emerged based on the core principle of dividing the total bandwidth into a large number of narrow, orthogonal subcarriers with distinct frequencies, rather than spreading a signal across the entire bandwidth.
These subcarriers of \ac{OFDM} are eigenfunctions of \ac{LTI} channels, enabling elegant per-subcarrier equalization when the channel remains approximately time-invariant over one symbol interval.
This condition was generally satisfied in the low-mobility, sub-6\,GHz operating regimes of \ac{4G}/\ac{5G}, and the subcarriers remain orthogonal and exhibit resilience to multipath fading with the use of a simple \ac{CP}.
However, they do not inherently provide frequency diversity, as each subcarrier is narrowband and may experience deep fading.

To remedy this, various techniques have been developed on top of the \ac{OFDM} paradigm:
\ac{mMIMO} beamforming providing channel hardening, intelligent resource scheduling, and powerful channel coding across subcarriers, all made practically feasible by the computationally efficient \ac{FFT}-based implementation of sinusoidal modulation.
Together, these enhancements allowed \ac{OFDM} to mature into a spectrally and hardware-efficient platform that has been the dominant wireless system design over the past two decades.

However, the nature of this \textit{robustness} deserves a closer examination.
Coded information can be distributed across subcarriers and \ac{OFDM} symbols, thereby providing frequency diversity when the allocation is sufficiently wide.
However, this type of diversity relies on the aforementioned \textit{external mechanisms} operating on top of narrowband subcarriers that do not themselves \textit{spread}, but rather map the coded information in a distributed fashion across non-spreading sinusoids.
Furthermore, while this mature technology and infrastructure continued to evolve, the time-invariance assumption of \ac{OFDM} was increasingly challenged by future application demands.

\vspace{-1ex}
\subsection*{\textbf{The Crisis (\acs{6G}) -- The Doppler Bottleneck}}

As we advance toward \ac{6G}, the time-invariance assumption underlying \ac{OFDM}'s success becomes increasingly difficult to maintain in many of the new envisioned use cases.
In high-mobility scenarios such as \ac{UAV}, \ac{V2X}, and \ac{LEO} satellite networks, the wireless channel becomes doubly dispersive, varying rapidly in both time and frequency.
This is further exacerbated by the planned use of higher-frequency bands in \ac{6G} including upper mid-band, \ac{mmWave}, and sub-\ac{THz} bands, where even moderate-to-high velocities (e.g., 120\,km/h in \ac{V2X}/\ac{UAV} at 71\,GHz) induce Doppler shifts corresponding to approximately $7\,\%$ of a typical 120\,kHz subcarrier spacing, thereby impairing orthogonality and causing detrimental \ac{ICI} in \ac{OFDM}-based systems.

As mentioned, there exist mitigation strategies within the \ac{OFDM} framework, but they require additional overhead and complexity, or are suitable only for specific operating regimes. 
For example, Doppler pre-compensation is effective for predictable, line-of-sight-dominated motion, such as single-\ac{LEO} orbital trajectories, but becomes challenging under unpredictable mobility with rich multipath scattering.
These approaches only fundamentally \textit{map} information across non-spreading subcarriers, rather than achieving intrinsic waveform-level \textit{spreading}.

This distinction has several critical consequences. 
First, coded diversity via distributed resource mapping becomes less effective for narrowband transmissions -- such as control signals, synchronization channels, and most importantly the small data packets that dominate practical traffic.
In such cases, allocations cannot span sufficiently distinct channel realizations, and the waveform itself must carry the required robustness.
Moreover, in a doubly dispersive channel, Doppler-induced \ac{ICI} does not create isolated fades but systematically degrades orthogonality across the entire grid, producing correlated impairments that frequency-domain interleaving and coding alone cannot efficiently resolve.

Furthermore, the emerging \acf{ISAC} paradigm reframes the concept of Doppler entirely \cite{Liu_JSAC22}: while communications has so far treated it as an impairment to be mitigated or compensated, sensing \textit{relies} on it as a primary observable quantity for estimating the velocity of environmental scatterers or targets.
A Doppler-fragile waveform is therefore less suitable for simultaneously supporting communication and sensing without resorting to slow-time approaches across multiple time slots, and waveform-level Doppler resilience is essential for the resource-efficient, multifunctional vision of \ac{6G+}.

\vspace{-1ex}
\subsection*{\textbf{The Synthesis (\acs{6G+}) -- The Resurrection of Spectrum Spreading}}

A promising waveform direction is therefore to employ subcarriers that inherently occupy an extended region of the \ac{TF} plane, thereby distributing each symbol's energy across \textit{both} time and frequency as an intrinsic structural property.

Therefore, waveform design must shift its primary focus back to robustness, and a form of spectrum spreading must return to the physical layer -- but this time not for multi-user interference suppression, as in \ac{3G}, nor for frequency diversity, which coded systems already achieve, but specifically for \textit{Doppler resilience}. 
However, at the same time, ever-increasing rate demands still remain a key priority, precluding a naive return to \ac{CDMA}-style frequency-domain wideband spreading. 
Instead, the \ac{6G+} paradigm calls for \textit{joint time-frequency spreading}, while preserving the multiplexing architecture, spectral efficiency, and maturity of the \ac{OFDM} legacy.

Two fundamentally different philosophies have emerged to realize this joint \ac{TF} spreading: \textit{2D isotropic spreading} via the \ac{DD} domain, exemplified by the \ac{OTFS} waveform family \cite{Hadani_WCNC17}, and \textit{sheared spreading} in the affine \ac{TF}-domain representation, where each symbol occupies a structured path across both time and frequency, exemplified by the \ac{AFDM} waveform family \cite{Bemani_TWC23}.
Both approaches have been shown to effectively achieve spreading across the \ac{TF} plane; however, they differ fundamentally in their mathematical formulation, architectural realization, and practical implications for the existing ecosystem and emerging \ac{6G} requirements.

\section*{The Return of Spreading -- But in What Form?}
\label{sec:return_of_spreading}

Throughout the dialectic rhythm above, the fundamental design choice at each generation has been the shape of the \textit{elementary per-symbol waveform} \rev{--} \rev{the shaped time-domain pulse of \ac{2G}, the \ac{PN}-multiplied wideband pulse of \ac{3G}, or the modulated sinusoid subcarrier of \ac{OFDM}.}
The same unit serves as the basic resource for multiple access, giving rise to \ac{TDMA}, \ac{CDMA}, \acs{OFDMA}, and their \ac{6G+} successors.

\begin{figure}[!b]
\vspace{-4ex}
\centering
\includegraphics[width=0.975\columnwidth]{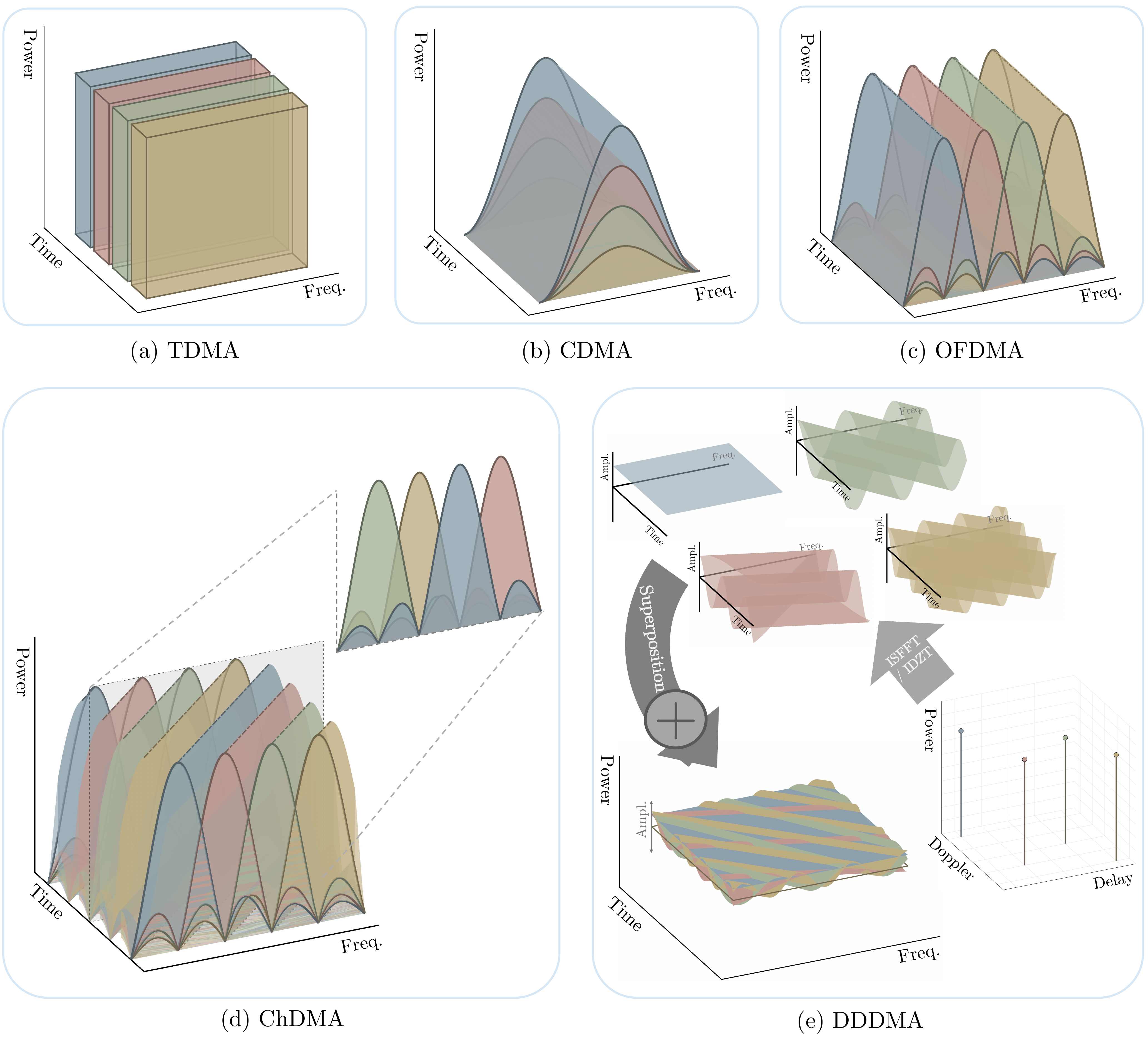}
\vspace{-1.5ex}
\caption{Five representative per-symbol waveform families and their footprints in the time-frequency plane, illustrating the progression toward the sheared and 2D isotropic footprints for doubly dispersive channels.
}
\label{fig:waveform_paradigms}
\end{figure}

As illustrated in Fig.~\ref{fig:waveform_paradigms}, the evolution of waveform spreading is intuitively captured by the footprint a single data symbol leaves on the \ac{TF} plane\rev{: each legacy scheme designs this footprint along only one dimension: a time slot, a code-spread wideband, or a narrowband subcarrier}.

However, the emerging requirements for joint \ac{TF} diversity necessitate a fundamental shift from these strictly 1D-localized footprints toward extended, 2D-coupled resource allocation. 
To ensure that each data symbol experiences the full doubly dispersive channel, two geometrically distinct paradigms have emerged: the \textit{sheared} approach illustrated as \ac{ChDMA} -- a term\rev{\footnote{\rev{The terms \acs{ChDMA} (and sheared spreading) and \acs{DDDMA} (and isotropic spreading) are descriptive labels coined internally to this article, and they are not established names in the literature.}}} reflecting the geometric shearing effect of the sinusoidal lines across the \ac{TF} plane as illustrated in Fig.~\ref{fig:waveform_paradigms}d -- structurally couples the two dimensions by employing an elementary chirp waveform that sweeps diagonally across the \ac{TF} plane; conversely, the 2D \textit{isotropic} approach illustrated as \ac{DDDMA} isotropically spreads a single \ac{DD}-domain symbol across the allocated \ac{TF} grid via a 2D transform, forming an equal-power 2D surface within the allocated \ac{TF} boundary of the signal as illustrated in Fig.~\ref{fig:waveform_paradigms}e.

Both approaches effectively distribute symbol energy across time and frequency, improving Doppler resilience and converting the channel into sparse, structured representations amenable to efficient estimation and detection.
\rev{Hence,} their differences lie less in achievable performance, with recent evaluations \rev{reporting comparable performance in high-mobility channels under matched detection and parametrization \cite{Zhang_CST26_UnifiedMulticarrierFramework, Rou_SPM24}}, but in realization and practicality, and the two approaches represent complementary yet competing design points along a common frontier.

\subsection*{The 2D Isotropic Spreading Paradigm: Strengths and Design Considerations}
\label{sec:2D_spreading_analysis}

The 2D isotropic spreading approach was pioneered by the original \ac{OTFS} proposal \cite{Hadani_WCNC17}, commonly referred to as \ac{MC}-\ac{OTFS} to distinguish it from its descendants, and has since matured into a rich family including Zak-\ac{OTFS} (OTFS 2.0) \cite{Mohammed_ZakOTFS_TIT25}, \rev{\ac{ODDM}, and more \cite{Zhang_CST26_UnifiedMulticarrierFramework}}.
Information symbols are placed in the \ac{DD} domain and spread across the \ac{TF} plane via a 2D transform, typically the \ac{ISFFT} paired with a Heisenberg transform, or the \ac{IDZT} for Zak-\ac{OTFS}, yielding the 2D footprint shown in Fig.~\ref{fig:waveform_paradigms}e.
By construction, each \ac{DD}-domain symbol \rev{spreads} across the entire allocated \ac{TF} grid -- the most literal realization of joint time-frequency spreading.
Indeed, the paradigm is the most mature representative of the spreading revival to date, supported by \rev{active theoretical work} and experimental demonstrations since 2017 \cite{Hadani_WCNC17}.

The \ac{DD}-domain construction offers two key advantages.
First, a doubly dispersive channel, which is rapidly time-varying in the \ac{TF} domain, maps to a nearly time-invariant and sparse representation in the \ac{DD} domain (see \cite[Fig.~1]{Rou_SPM24}).
To appreciate this sparsity, consider a \ac{V2X} scenario with a small number of dominant environment scatterers -- a preceding vehicle, a roadside building, and an overpass -- each giving rise to a single resolvable path with a distinct delay and Doppler shift determined by its range and relative velocity. 
In the \ac{TF} domain, these paths combine into a rapidly fluctuating, dense channel across the entire grid, whereas in the \ac{DD} domain, the same channel reduces to just a handful of isolated point-like taps, one per scatterer, whose coordinates in the \ac{DD} grid correspond to its path delay and Doppler shift, making estimation and equalization structurally simple.
Second, the representation is intrinsically sensing-friendly, as \ac{DD} coordinates translate directly into range and radial velocity, making \ac{OTFS} a natural fit for \ac{ISAC}.
Furthermore, the \ac{DD}-domain input/output relation becomes predictable and nonfading, and even extractable from a single pilot symbol.

However, these strengths come with some design considerations relative to \ac{OFDM} systems.
The first concerns pulse shaping. 
By the Balian-Low theorem, any pulse forming a valid basis on the critically sampled \ac{TF} Heisenberg lattice is poorly localized in time, frequency, or both, making a jointly well-localized pulse physically unrealizable.
Consequently, rectangular pulses commonly assumed in \ac{MC}-\ac{OTFS} analyses sacrifice frequency localization and produce \ac{DD}-domain spreading and leakage under fractional delay and Doppler.

Zak-\ac{OTFS} \cite{Mohammed_ZakOTFS_TIT25} addresses this limitation through the crystallization condition, which guarantees a predictable and non-fading \ac{DD}-domain I/O relation when the delay and Doppler periods exceed the effective channel spreads, restoring well-defined behavior under practical pulses. 
However, achieving improved \ac{DD}-domain localization in Zak-\ac{OTFS} explicitly requires structural machinery that is absent from an \ac{OFDM}-like pipeline, such as factorizable \ac{DD}-domain pulse filters whose transmit and receive operations involve convolutions, along with the associated time and bandwidth expansion overhead.

These considerations do not diminish the technical maturity or effectiveness of the 2D isotropic paradigm; rather, they reflect the engineering depth invested in making \ac{DD}-domain modulation practical and highly effective.

They do motivate a natural question, however: can the structural benefits of joint \ac{TF} spreading -- sparse effective channel, full \ac{TF} diversity, and intrinsic Doppler resilience -- be obtained through a construction that stays closer to the \ac{OFDM} legacy and intuition?

\section*{From Sinusoids to Chirps: The Sheared Spreading Alternative}
\label{sec:afdm}

Motivated by the above question, we argue that the shearing approach offers a compelling synthesis, exemplified by the \ac{AFDM} family, which retains the structural benefits of joint \ac{TF} spreading of the 2D isotropic paradigm, while maintaining maximal backward compatibility with the mature \ac{OFDM} technology, and introduces a unique parametrizable degree of freedom.

\rev{In light of this, extensive surveys have rigorously investigated and benchmarked the mature 2D isotropic paradigm against the alternative paradigms through systematic, KPI-based comparisons spanning the vast family of next-generation multicarrier waveforms, including chirp-based approaches such as \ac{OCDM} and \ac{AFDM} \cite{Zhang_CST26_UnifiedMulticarrierFramework, Rou_SPM24}.
With the comparative ground already thus well covered, we clarify that this article is not intended to be another comparison, but a perspective piece: a historical-architectural synthesis of how spectrum spreading returns in \ac{6G+}, coupled with a forward-looking assessment of the sheared-spreading paradigm aimed at motivating the community toward the open problems this evolution exposes.}

\rev{Nonetheless, for completeness and the convenience of the readers, we also include at the end of the article a compiled summary of the comparative results and literature for the prime candidates in Table~\ref{tab:comparison}, before focusing on the chirp-based paradigm in the following sections.}

\subsection*{Chirps as a Generalized Sinusoid}
\label{sec:chirp_vs_sinusoid}

\begin{figure}[h]
\vspace{1ex}
\centering
\includegraphics[width=1\columnwidth]{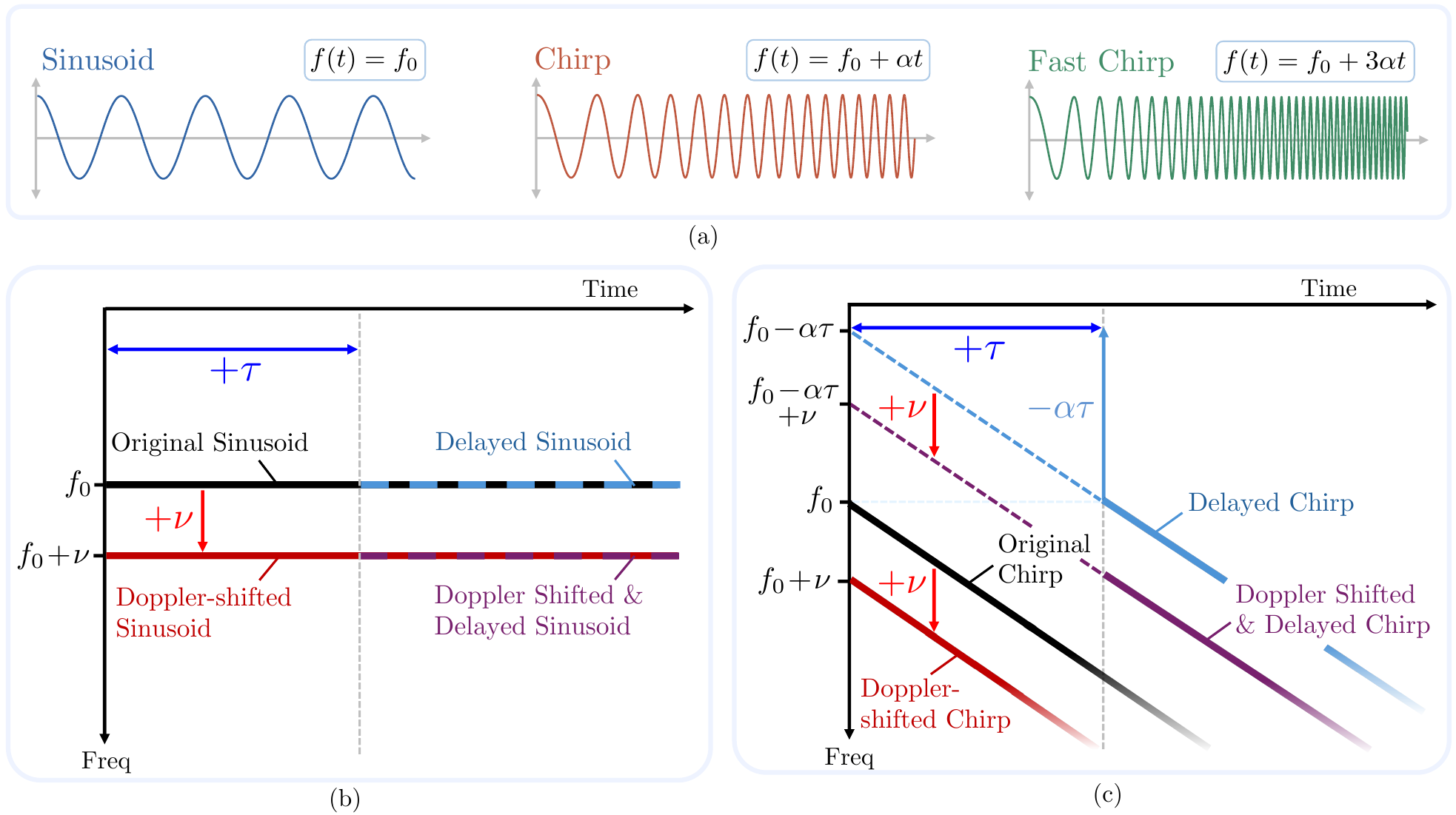}
\vspace{-4ex}
\caption{Illustration of sinusoid and chirp behaviors: 
(a) time-domain waveforms, highlighting the sinusoid as the special case of zero chirp rate ($\alpha = 0$), and two exemplary chirps of different rates; 
(b) instantaneous frequency versus time for sinusoids under delay and Doppler, showing that a delay $\tau$ preserves the frequency $f_0$ while a Doppler shift $\nu$ maps the subcarrier to a distinct sinusoid at $f_0 + \nu$; 
and (c) instantaneous frequency versus time for chirps under delay and Doppler, showing that both effects act on the starting frequency alone, yielding a net displacement of $\nu - \alpha\tau$ while the chirp rate $\alpha$ and waveform shape are fully preserved.
Unlike the sinusoid, which becomes a fundamentally different signal under Doppler, the chirp remains in the same waveform family under both delay and Doppler, becoming merely a shifted version of itself.}
\label{fig:sinusoid_vs_chirp}
\end{figure}

A \textit{chirp} is any signal with a time-varying instantaneous frequency, with the general class including linear, quadratic, hyperbolic, and exponential variants.
In the wireless literature, however, ``chirp'' is conventionally used to denote the \textit{linear chirp}, i.e., a linear frequency-modulated signal with instantaneous frequency $f(t) = f_0 + \alpha t$ and corresponding quadratic phase, parametrized by the \textit{chirp rate} $\alpha$.
Under this definition, the sinusoid is recovered as the special case $\alpha = 0$, as illustrated in Fig.~\ref{fig:sinusoid_vs_chirp}a.
In other words, the relationship between the sinusoid and the linear chirp is one of generalization rather than substitution:
introducing a chirp rate adds a new degree of freedom without discarding any property of the sinusoidal baseline.

The motivation for shifting from sinusoids to chirps is primarily channel-theoretic. 
Sinusoids are the eigenfunctions of \ac{LTI} channels, which is why \ac{FFT}-based \ac{OFDM} diagonalizes the channel and achieves subcarrier orthogonality in static environments \cite{Nee_OFDM2000}. 
However, when the channel becomes \ac{LTV} -- i.e., doubly dispersive -- sinusoids lose this eigenfunction property, and finding a universal orthonormal basis that exactly diagonalizes general \ac{LTV} channels is non-trivial.
Therefore, rather than seeking a strict eigenbasis, the practical strategy is to project the signal onto a basis that induces a structured, sparse representation of the channel, exploiting the compactness of the delay-Doppler spreading function. 
In this context, chirps serve as a natural generalization of sinusoids: by matching the chirp parameters to the linear delay-Doppler coupling of the propagation paths, the doubly dispersive channel is converted into a sparse, quasi-diagonal effective channel matrix in the \ac{DAFT} domain, making chirps a suitable basis when time invariance is lost \cite{Bemani_TWC23}.

Geometrically, the distinction is immediate on the \ac{TF} plane (Fig.~\ref{fig:sinusoid_vs_chirp}): a sinusoid occupies a horizontal line at fixed frequency $f_0$, whereas a chirp traces a diagonal line with slope $\alpha$.
As illustrated in Fig.~\ref{fig:sinusoid_vs_chirp}b and Fig.~\ref{fig:sinusoid_vs_chirp}c, the two waveforms respond to delay and Doppler in fundamentally different ways.
For a sinusoid, delay and Doppler act on \textit{different} signal properties: a delay $\tau$ produces a phase rotation proportional to both the delay $\tau$ and the carrier frequency $f_0$ that leaves the waveform itself (sinusoid at frequency $f_0$) unchanged, while a Doppler shift $\nu$ translates the signal to a different sinusoid centered at frequency $f_0 + \nu$.

For a chirp, by contrast, both delay and Doppler act on the \textit{same} property, namely the starting frequency: delay produces a shift of $-\alpha\tau$~Hz, Doppler produces a shift of $+\nu$~Hz, and the net effect is a single displacement $\nu - \alpha\tau$~Hz in the starting frequency while the chirp rate $\alpha$ is preserved.
The chirp therefore retains its key functional parameter, becoming merely a shifted version of itself in the \ac{TF} plane -- analogous to how a sinusoid is preserved under delay alone.

The structural significance of this is the collapse of the two-dimensional delay-Doppler effect into a single scalar displacement $\nu - \alpha\tau$ on a preserved waveform family, with $\alpha$ acting as a tunable coupling coefficient between delay and Doppler.
It should be clarified that this does \textit{not} mean that shifted chirp subcarriers never overlap with other chirps in the \ac{TF} plane, rather, it implies that the chirp parameter $\alpha$ can be designed to control where the multipath contributions corresponding to each delay-Doppler pair land on the frequency axis.
Indeed, by selecting $\alpha$ according to the full-diversity condition \cite{Bemani_TWC23}, the net displacements of all distinct propagation paths in a doubly dispersive channel are guaranteed to be mutually distinct in the affine-frequency (\ac{DAFT}) domain, and hence in the effective channel.
This converts the dense, unstructured \ac{ICI} of sinusoidal subcarriers -- which arises because the uncontrolled Doppler shifts of multiple paths cannot be jointly resolved on a fixed-frequency grid -- into a well-organized, deterministic structure that enables effective channel estimation and equalization.

Importantly, chirp-based signal processing is not new to wireless systems. 
They have long been the foundation of \ac{FMCW} radar, where \rev{linear frequency sweeps enable} range and velocity estimation through \rev{simple} matched-filter \rev{(de-chirping) operations}.
Beyond sensing, chirps have also found natural roles in communications: they underpin \ac{LoRa}, where binary-orthogonal up- and down-chirps enable long-range, low-power \ac{IoT} connectivity, and form the subcarrier set of \ac{OCDM} \cite{Ouyang_OCDM15}, which spreads each symbol across orthogonal chirp subcarriers spanning the full bandwidth once.

\vspace{2ex}
\subsection*{Affine Frequency Division Multiplexing (AFDM): The Flexible Chirp-Based Multicarrier}
\label{sec:afdm_architecture}

Following this line of development, \ac{AFDM} builds a multicarrier waveform from the chirp subcarriers introduced above via the \ac{DAFT} \cite{Bemani_TWC23}, a generalized \ac{FFT} parametrized by two chirp rates, customarily denoted $c_1$ and $c_2$.
The \ac{DAFT} also reduces to standard transforms at specific parameter settings: $(c_1, c_2) = (0, 0)$ recovers the \ac{FFT} of \ac{OFDM}, while $(c_1, c_2) = (\frac{1}{2N}, \frac{1}{2N})$, with $N$ denoting the number of subcarriers, yields the \ac{DFnT} of \ac{OCDM}.
Consequently, \ac{AFDM} includes \ac{OFDM} and \ac{OCDM} as special cases. 
This enables a single, unified framework in which transitioning between waveform modes requires only a simple parameter adjustment, analogously to adaptive modulation and coding in response to channel conditions and application demands \cite{Yin_AFDM25} (i.e., $c_1 = 0$ for static channels, $c_1 > 0$ for mobile channels).

The \ac{DAFT}-based approach defines $(c_1, c_2)$ as a highly flexible two-parameter design space, in which the two parameters play asymmetric roles:
The first parameter, $c_1$, defines the chirp rate controlling the slope of the chirps in the \ac{TF} plane -- or the amount of shearing of the subcarriers in Fig.~\ref{fig:waveform_paradigms} -- and hence the extent of \ac{TF} spreading.
When $c_1 = \frac{1}{2N}$, as in the \ac{OCDM} case, the chirp subcarriers sweep the entire bandwidth exactly once over the signal duration, and when $c_1 > \frac{1}{2N}$, they sweep the bandwidth multiple times by wrapping around frequency boundaries. 

In \ac{AFDM}, the \rev{condition on} $c_1$ that ensures full diversity over the doubly-dispersive channel is \rev{a bound determined from} the number of subcarriers and the channel's maximum normalized Doppler shift \cite{Bemani_TWC23}.
\rev{Consequently, a single cell-wide $c_1$ dimensioned for the worst-case served Doppler preserves full diversity for every heterogeneous user within the cell's delay-Doppler profile, with lower-mobility users retaining a wider separation margin.}

Under this condition, each unique propagation path contributes a single well-separated component to the effective channel matrix in the affine-frequency domain.
\rev{It is not the chirps alone, however, that enable the robustness to Doppler: the full diversity enabled by sheared spreading is completed and harvested only with the dedicated support of novel \ac{DAFT}-domain signal processing techniques including channel estimation and equalization, which have been richly developed and growing \cite{Yin_AFDM25, Zhang_CST26_UnifiedMulticarrierFramework, Bemani_TWC23, Rou_AFDM6G}.
}

On the other hand, the second chirp parameter, $c_2$, does not affect subcarrier orthogonality or the diversity condition, and instead controls the effective precoder that rotates the phases of the symbols before chirp-subcarrier modulation.
Therefore, this flexibility of $c_2$ can be exploited for auxiliary design objectives, all without \rev{affecting} the aforementioned diversity \rev{condition on the chirp rate $c_1$}.
\rev{This} unique degree of freedom \rev{is the key to enabling multifunctional applications}, such as \ac{AF} shaping for sensing, spectral shaping for \ac{PAPR} and \ac{OOBE} reduction, \ac{IM}, and \ac{PLS} \cite{Rou_AFDM6G, Yin_AFDM25}.

In all, we highlight that through \ac{AFDM}, the system can achieve \textit{parametrizable} sheared spreading of the subcarriers and symbols.

\subsection*{The Practical Case for Parametrizable Sheared Spreading}
\label{sec:chirps_vs_dd}

As established in the discussion of the 2D isotropic paradigm, both approaches achieve the fundamental goal of joint \ac{TF} spreading and offer analogous structural benefits in doubly dispersive channels. 
Therefore, their differences lie more in architectural realization and practical implications than in achievable performance. 

\begin{figure}[h]
\centering
\vspace{-1ex}
\includegraphics[width=0.875\columnwidth]{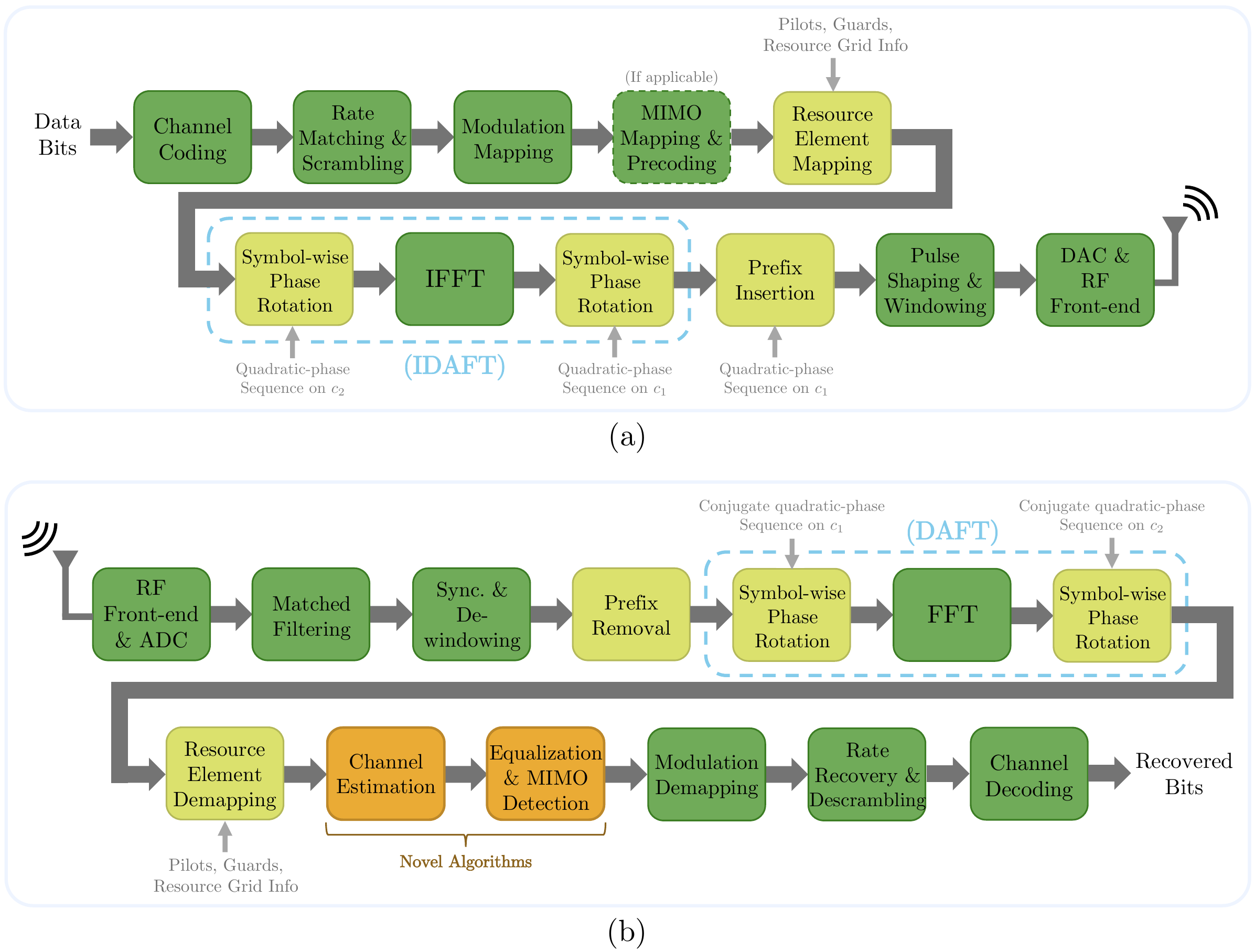}
\vspace{-2ex}
\caption{An example of an \ac{AFDM} end-to-end transceiver architecture, illustrating both (a) the transmitter and (b) the receiver pipelines. 
As illustrated in green blocks, \ac{AFDM} directly reuses most \ac{OFDM} blocks.
The yellow blocks represent a ``soft update,'' meaning they rely on existing \ac{OFDM} architecture but require the changing of parameters (i.e., applying specific chirp-phase sequences or \ac{CPP} masks). 
Finally, the orange blocks denote areas that require algorithmic novelty -- in digital baseband processing and not new hardware -- to execute \ac{AFDM}-specific channel estimation and joint equalization and \ac{MIMO} detection.}
\label{fig:transceiver_pipeline}
\vspace{-8ex}
\end{figure}

The \ac{DAFT} structure admits a simple transmitter decomposition that remains close to the conventional \ac{OFDM} pipeline. 
The \ac{IDAFT} is equivalent to a sample-wise multiplication with a quadratic-phase vector with chirp rate $c_2$, followed by a standard \ac{IFFT} and second sample-wise multiplication with another quadratic-phase vector with chirp rate $c_1$, as illustrated in Fig.~\ref{fig:transceiver_pipeline}. 
Therefore, any \ac{FFT}-based baseband designed for \ac{OFDM} can be extended to \ac{AFDM} by inserting two element-wise, unit-magnitude phase rotations around the existing core, adding only a computational overhead \rev{of $2N$ unit-magnitude phase rotations,} on top of the unchanged $\mathcal{O}(N\log N)$ \ac{FFT} operation.

\clearpage

\rev{Notably,} this implementation uses exact element-wise multiplications to modulate the information symbols onto chirps, requiring no idealized assumptions in the \ac{DAFT} definition \rev{itself and no} supplementary compensation filtering for time-frequency pulse shaping.
More importantly, this element-wise phase-rotation mechanism is already present in various forms across deployed \ac{OFDM} frameworks including \ac{LTE}, \ac{5G}, and IEEE 802.11 standards; the two chirp multiplications required for \ac{AFDM} thus represent a physically grounded extension of an operation class that existing \ac{OFDM} hardware and firmware already support.

At the receiver, \ac{AFDM} requires a \ac{CPP}, which is a \ac{CP} with a deterministic phase rotation per sample \cite{Bemani_TWC23, Rou_SPM24}.
\rev{Interestingly though, \ac{CPP} reduces to a plain \ac{CP} whenever $2Nc_1$ is an integer and $N$ is even, so prefix dimensioning can easily follow \ac{OFDM} practice.}
The \ac{CPP} preserves circular convolution and yields a structured, effective channel in the affine-frequency domain, enabling low-complexity detection algorithms, including banded-\ac{MMSE} and sparse message-passing variants \rev{-- though, as with \ac{OTFS}, the one-tap equalization of \ac{OFDM} is forfeited, since the effective channel is not diagonal.}

\rev{The corresponding channel estimation uses a single-pilot scheme with 1D padding in the affine-frequency domain \cite{Bemani_TWC23}, and such components can be integrated into an \ac{OFDM}-style transceiver by reframing the processing frame from 1D-frequency to 1D-affine-frequency, leaving analog front-ends untouched.
This ecosystem continuity is the primary practical advantage of sheared spreading over a frequency-domain resource rather than a full 2D \ac{DD} grid: the \ac{FFT}-core numerology remains intact for spacing, prefixing, and pilots, and the \textit{subcarriers} remain the natural multiple-access resource, carrying legacy \acs{OFDMA} logic directly to \acs{AFDMA}.}

\vspace{3ex}
\section*{The Chirp-Based Research Landscape and the Road Ahead}

\begin{landscape}
\begin{figure}[H]
\centering
{\includegraphics[width=1\columnwidth, height=0.5775\columnwidth]{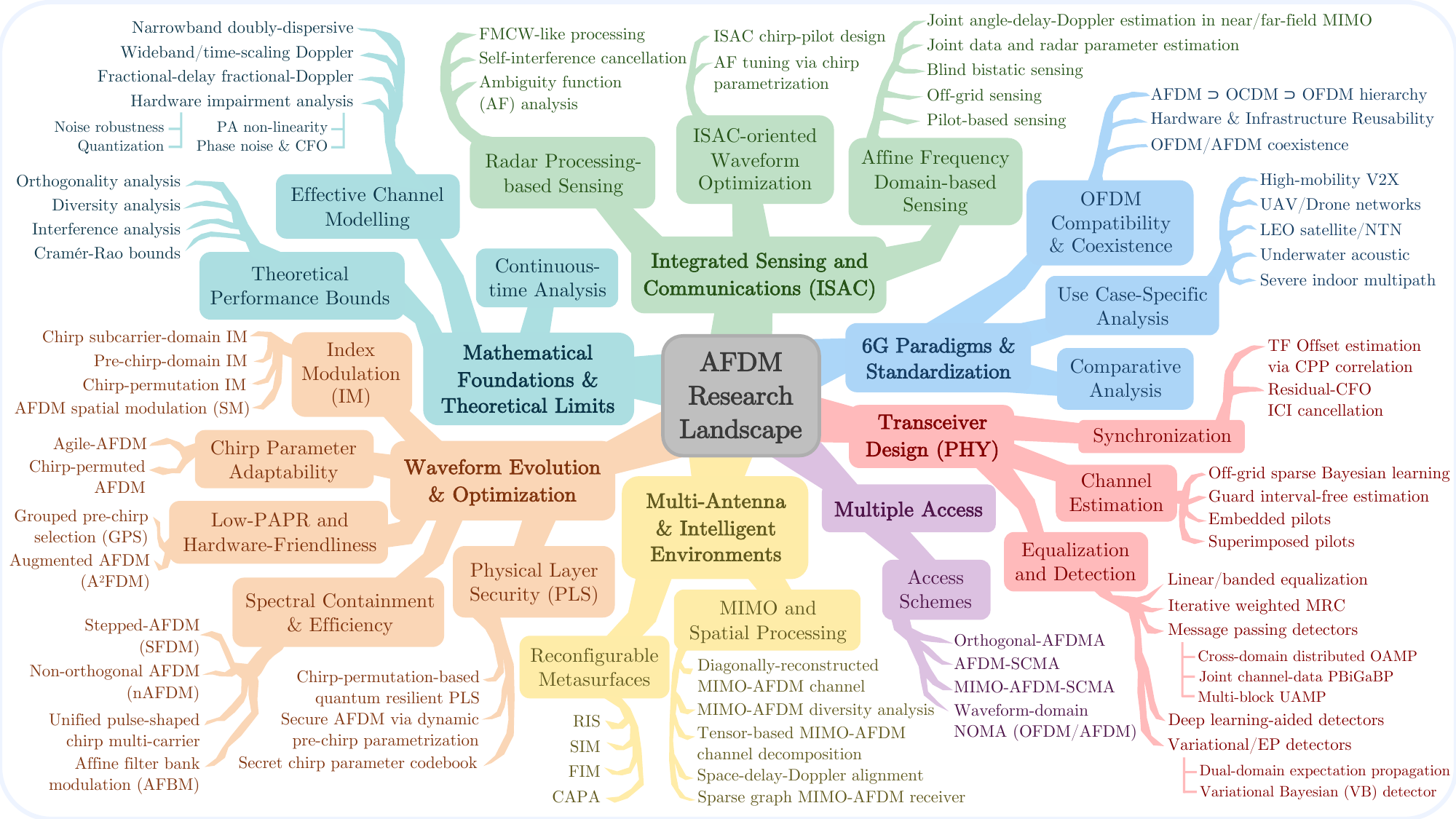}%
  }
\vspace{-4ex}
\caption{A representative mind map of the \ac{AFDM} research landscape that has emerged since the waveform's formal introduction in 2021, organized around seven primary research axes: mathematical foundations and theoretical limits, MIMO processing and intelligent environments, \ac{6G} paradigms and standardization relevance, transceiver signal processing, \ac{ISAC}, multiple access, and waveform evolution and optimization. 
Each primary axis branches into sub-topics covering prominent contributions, and some exemplary techniques and contributions when applicable.
Note that this figure is intended as a structured overview of representative and prominent topics and results across the landscape and is not exhaustive; the full body of published contributions is considerably larger and continues to grow.}
\label{fig:afdm_landscape}
\end{figure}
\end{landscape}

Since its introduction in 2021, \ac{AFDM} research has expanded across nearly every axis of physical-layer design, \rev{extending} the parametrizable chirp structure well beyond \rev{its original Doppler-resilience objective}.
\rev{As illustrated by the the vast landscape in Fig.~\ref{fig:afdm_landscape}, the research spans transceiver design through waveform-domain optimization, exploiting a chirp-parameter space unusually small and tractable -- an attractive target for \acs{AI}-native waveform adaptation.}

\rev{Amongst these, \ac{ISAC} is the sharpest case: the native chirp structure enables \ac{FMCW}-like processing to extract range and velocity from a single symbol, rather than multiple symbols per coherent processing interval as in \ac{OFDM} radar \cite{Zhu_TWC26_AFDMISACFMCWPerspective}.
Moreover, the full-diversity condition renders each path's delay-Doppler pair uniquely identifiable in the \ac{DAFT} domain -- a property shared with \ac{OTFS}, whose delay-Doppler native geometry likewise renders channel estimation and radar parameter estimation equivalent \cite{Rou_SPM24}.}

\vspace{0.5ex}
\subsection*{\rev{Principal open issues}}

\rev{Nonetheless, several open issues remain before \ac{AFDM} can be considered a mature \ac{6G} candidate.
Some advantages are so far only conceptually identified and require deeper technical treatment, while the most pressing challenges are integrative in nature, recently surfacing in \acs{3GPP} RAN1 feature-lead summaries \cite{Rou_AFDM6G, Yin_AFDM25}.
Chirp-parameter signaling, for instance, is non-trivial under per-\acs{RB} and per-\acs{BWP} requirements, as is synchronization at high Doppler, where conventional methods may not directly apply and the chirp structure itself may offer the solution \cite{Luo_arXiv26_TowardsStandardizingAFDM}.
\ac{OFDM} coexistence is also an under-addressed yet important area: as a direct generalization of \ac{OFDM}, \ac{AFDM} can in principle be dimensioned within legacy numerology to enable efficient multi-user and even multi-generation coexistence.}
\rev{Spectral containment also requires care: because every \ac{DAFT} symbol spans the full band, the guard band cannot be formed simply by nulling the outer subcarriers, and hence requires either a dedicated filtering stage or an alternate mode in which the chirps are formulated as a precoded \ac{OFDM} via cascading the \ac{IDAFT} ahead of the \acs{NR} \ac{CP}-\ac{OFDM} chain and mapping onto active \acs{RB}s \cite{Luo_arXiv26_TowardsStandardizingAFDM}.}
\rev{Testbed validation also remain limited, where over-the-air demonstrations have only recently emerged, and trials at standardized numerology, especially under the envisioned use-case scenarios, are scarce \cite{Rou_AFDM6G}.}

\rev{\vspace{2ex}
\section*{Conclusion}

In this article, we have provided the perspective that the evolution of the wireless physical layer is better read as a pendulum between rate and robustness than a monotonic climb in rate, and \ac{6G+} is where it swings back -- and spreading must return, now jointly across the \ac{TF} plane, for Doppler resilience.
The chirp-based alternative generalizes rather than replaces -- \ac{AFDM} contains \ac{OFDM} and \ac{OCDM} as parameter settings, collapses delay-Doppler into a single displacement of a preserved waveform, and reduces to two phase rotations around an unchanged \ac{FFT} core, while delivering full diversity, single-symbol \ac{ISAC}, and a flexible parameter for secondary functionalities and objectives.
The move from sinusoids to chirps is thus technically motivated and evolutionarily economical, and what remains is integrative -- signaling, coexistence, and over-the-air validation will decide the case.}

\begin{landscape}
\begin{table}[H]
\centering
\newcommand{\tabFontSize}{\scriptsize}  %
\newcommand{\tabCapSize}{}              %
\newcommand{\tabLineSpread}{1}       %
\newcommand{\tabRowStretch}{0.95}       %
\newcommand{\tabColSep}{3pt}            %
\newcommand{\tabTotalWidth}{21.9cm}     %
\newcommand{\wLabel}{0.1324}            %
\newcommand{\wOFDM} {0.1895}            %
\newcommand{\wOCDM} {0.1895}            %
\newcommand{\wOTFS} {0.2443}            %
\newcommand{\wAFDM} {0.2443}            %
\newcommand{\tabRuleWidth}{0.4pt}       %
\newcommand{\tabDoubleRuleSep}{2pt}     %
\newcommand{\tabHyphenPenalty}{0}       %
\newcommand{\tabTolerance}{9999}        %
\newcommand{\tabMarkColor}{black}     %
\ifdefined\tabTW\else\newlength{\tabTW}\fi
\setlength{\tabTW}{\tabTotalWidth}
\colorlet{tabmark}{\tabMarkColor}
\newcommand{\rv}[1]{{\color{tabmark}#1}}
\newcolumntype{K}[1]{>{\raggedright\arraybackslash\hspace{0pt}\color{tabmark}}p{#1\tabTW}}
\caption{\tabCapSize\rv{Comparative characteristics of key performance indicators (KPIs) between the 2D isotropic (OTFS-family) and parametrized sheared (AFDM-family) spreading paradigms, against the OFDM and OCDM baselines. OCDM belongs to the chirp-based sheared family, with chirp parameters fixed to $c_1{=}c_2{=}\frac{1}{2N}$, but is kept as its own column to isolate the effect of a fixed chirp rate against the parametrizable chirps of AFDM. The representative high-mobility operating point for evaluations follow \cite[Sec.~VII]{Bemani_TWC23} (QPSK, $N=256$, max. normalized Doppler $\alpha_{\max}=2$, e.g., $540$\,km/h at $4$\,GHz). Entries without citations state a general characteristic, or results whose source could not be cited within the reference limit.}}
\label{tab:comparison}
\vspace{-1ex}
\tabFontSize
\setstretch{\tabLineSpread}
\renewcommand{\arraystretch}{\tabRowStretch}
\setlength{\tabcolsep}{\tabColSep}
\setlength{\arrayrulewidth}{\tabRuleWidth}
\setlength{\doublerulesep}{\tabDoubleRuleSep}
\hyphenpenalty=\tabHyphenPenalty \exhyphenpenalty=\tabHyphenPenalty
\pretolerance=-1 \tolerance=\tabTolerance
\begin{tabular}{|K{\wLabel}||K{\wOFDM}|K{\wOCDM}|K{\wOTFS}|K{\wAFDM}|}
\hline
\textbf{Characteristic} & \textbf{OFDM} & \textbf{OCDM} & \textbf{OTFS} & \textbf{AFDM} \\
\hline\hline
Spreading geometry $~$ (TF footprint)
& Narrowband subcarriers, no TF spreading at the waveform level \cite{Yin_AFDM25}
& Exactly single sweep (sheared spreading) over full TF
& Two-dimensional isotropic spreading over full TF
& Parametrizable number of sweeps (sheared spreading) over full TF \cite{Yin_AFDM25, Luo_arXiv26_TowardsStandardizingAFDM} \\
\hline
Receiver complexity and architecture
& One-tap equalization in LTI; ICI under Doppler \cite{Zhang_CST26_UnifiedMulticarrierFramework}
& Not one-tap under Doppler: block LMMSE in LTV; single-tap FDE only if quasi-static \cite{Ouyang_OCDM15}
& Not one-tap: block MMSE \cite{Mohammed_ZakOTFS_TIT25} or iterative detectors
& Not one-tap: banded MMSE \cite{Bemani_TWC23} or iterative detectors \\
\hline
Spectral efficiency (guard overhead)
& Per-symbol cyclic prefix (CP), overhead high at short symbols \cite{Bemani_TWC23}
& OFDM-like (per-symbol CP), identical block/guard structure \cite{Ouyang_OCDM15}
& Single guard per frame; reduced-CP and Zak variants available
& Per-symbol chirp-periodic prefix (CPP), OFDM-like cost and structure \cite{Bemani_TWC23} \\
\hline
Pilot overhead
& Comb-type/scattered (DMRS), interpolated
& OFDM channel-estimation schemes adaptable \cite{Ouyang_OCDM15}
& Embedded pilot with 2D guard region in DD domain
& Embedded pilot with 1D guard region in DAFT domain \cite{Bemani_TWC23} \\
\hline
Peak-to-average-power ratio (PAPR) and out-of-band emission (OOBE)
& High PAPR, mature reduction techniques available; high OOBE, countered by windowing/filtering
& High PAPR and OOBE(OFDM-comparable); required dedicated spectral shaping \cite{Ouyang_OCDM15}
& Lower PAPR from DFT-based spreading of symbol power \cite{Rou_SPM24}; OOBE highest under rectangular pulses, contained by SRRC/ODDM
& High PAPR, statistically OFDM-comparable \cite{Zhang_CST26_UnifiedMulticarrierFramework}; reduction available via chirp parametrization, and dedicated spectral shaping required \\
\hline
Related waveforms and variants
& DFT-s-OFDM; filtered OFDM (F-OFDM); FBMC, UFMC, GFDM
& \textit{OCDM is itself a special case of AFDM} \cite{Rou_AFDM6G}
& Zak-OTFS; ODDM; OTSM; ODSS; DDAM; reduced-CP variants
& Chirp-permuted AFDM; DAFT-s-AFDM; AFBM; Non-orthogonal AFDM, SFDM \\
\hline
Diversity order in doubly-dispersive (LTV) channels
& Order one; cannot separate paths \cite{Bemani_TWC23}
& Order one; better path separation than OFDM, but paths can overlap \cite{Bemani_TWC23}
& Asymptotically order one without precoding, but full-diversity with precoding; Zak-OTFS full-diversity under crystallization \cite{Mohammed_ZakOTFS_TIT25}
& Full-diversity for chirp rates satisfying the closed-form $c_1$ condition \cite{Bemani_TWC23} \\
\hline
Fractional delay--Doppler behavior
& No DD grid, so fractional delay--Doppler enters only as ICI; paths remain unseparated \cite{Bemani_TWC23}
& Order-one diversity as OFDM, overlapping paths can add destructively \cite{Bemani_TWC23};
& Fractional-Doppler spreading across all Doppler indices across the block-grid \cite{Rou_SPM24}; Zak-OTFS crystallization mitigates
& Leakage confined by widened guards via the slack parameter $\xi_\nu$ \cite{Bemani_TWC23}; not immune, but can be controlled to maintain banded structure \cite{Rou_SPM24}\\
\hline
MIMO integration and maturity
& Mature MIMO-OFDM in 5G NR
& Isolated MIMO-OCDM sensing studies, comparable to MIMO-OFDM;
& DD-domain interference managed globally; pilot/guard/estimation cost grows with antenna count
& Full diversity proven \cite{Yin_AFDM25}; channel estimation becomes multi-dimensional \cite{Luo_arXiv26_TowardsStandardizingAFDM}; robust under hardware impairments \\
\hline
Channel estimation
& Frequency domain over per-RE pilots (DMRS), interpolated; only diagonal acquired, ICI unestimated
& Same frequency-domain CFR as OFDM \cite{Ouyang_OCDM15}; no single chirp-domain-pilot as path coincide \cite{Bemani_TWC23}
& DD-domain, single embedded pilot with 2D guard, full DD representation \cite{Bemani_TWC23}; single-pilot acquisition under crystallization \cite{Mohammed_ZakOTFS_TIT25}
& DAFT-domain, embedded single pilot with 1D guard, full DD representation  under optimal chirp parametrization \cite{Bemani_TWC23, Rou_AFDM6G} \\
\hline
Synchronization
& Mature; decades deployed \cite{Zhang_CST26_UnifiedMulticarrierFramework,Nee_OFDM2000}
& No dedicated study; sync explicitly left open in the OFDM-compatibility analysis \cite{Ouyang_OCDM15}
& Open; can use OFDM-like synchronization, and DD-domain approaches actively researched and proposed
& Open; can use OFDM-like synchronization; chirp sequences already underpin \rv{LTE sync and NR random-access preambles} \cite{Luo_arXiv26_TowardsStandardizingAFDM} \\
\hline
RF/Hardware impairments
& PN/PA dominant, countered by pre-distortion, phase tracking, etc. \cite{Zhang_CST26_UnifiedMulticarrierFramework}
& Rated intermediate in KPI survey \cite{Zhang_CST26_UnifiedMulticarrierFramework}; \rv{no dedicated study identified}
& Proven resiliency to PN/CFO, up to an order of magnitude lower BER than OFDM \cite{Zhang_CST26_UnifiedMulticarrierFramework}
& Proven waveform-level resiliency to PN/CFO, outperforms OFDM in BER \cite{Zhang_CST26_UnifiedMulticarrierFramework}\\
\hline
Multiple access
& OFDMA: standardized orthogonal subcarrier allocation
& One uplink design study, inter-user interference analyzed \rv{for frequency-selective channels only}
& Studied MU-MIMO-OTFS, NOMA, RSMA, inter-user interference harder than OFDMA
& Studied AFDMA and AFDM-SCMA \cite{Yin_AFDM25}; native DAFT-s-AFDMA in the NR uplink \cite{Luo_arXiv26_TowardsStandardizingAFDM} \\
\hline
ISAC / sensing readiness
& Mature radar processing, Doppler over slow-time only \cite{Zhu_TWC26_AFDMISACFMCWPerspective}, lowest average ranging sidelobe
& OFDM-equivalent detection under MF/GLRT, comparable AF; parameter estimation unaddressed
& Native DD-domain sensing, FMCW-accurate at full rate, as is OFDM over slow time; estimation by iterative approximated-ML search
& FMCW-equivalent subcarriers with chirp-periodic $c_1$ \cite{Zhu_TWC26_AFDMISACFMCWPerspective}, single-symbol range/velocity without full-duplex SIC; chirp-rate tunable AF; sensing near the CRLB \\
\hline
Standardization maturity
& 3GPP baseline (4G \~ 5G NR); CP-OFDM / DFT-s-OFDM reaffirmed as the 6G baseline
& None; a special case of AFDM \cite{Rou_AFDM6G}
& Early standardization efforts; not adopted, but identified as potential candidate
& Early standardization efforts; promising on OFDM-compatibility grounds \cite{Rou_AFDM6G};  identified as potential candidate \\
\hline
\end{tabular}
\end{table}
\end{landscape}

\newpage


\section*{Biographies}

\noindent\textbf{Hyeon Seok Rou} (Member, IEEE) received the B.Sc. degree from Jacobs University Bremen, Germany, in 2021, and the Ph.D. degree from Constructor University Bremen, Germany, in 2024. He is currently a Research Assistant Professor with the Korea Advanced Institute of Science and Technology (KAIST), Daejeon, South Korea, under the Brain Pool Program of the National Research Foundation of Korea. His research interests include affine frequency division multiplexing (AFDM), integrated sensing and communications, and over-the-air computing. He has published over 50 peer-reviewed papers and is a coinventor of over five granted patents.\par\medskip

\noindent\textbf{Giuseppe Thadeu Freitas de Abreu} (Senior Member, IEEE) received the master's and Ph.D. degrees from the Yokohama National University, Japan. He is a Professor of electrical engineering with the School of Computer Science and Engineering, Constructor University, Bremen, Germany. His research interests include communications and signal processing, quantum computing, and metasurfaces. He published over 400 papers, coauthored over 70 patents, and cofounded 2 startup companies.\par\medskip

\noindent\textbf{Emil Bj\"{o}rnson} (Fellow, IEEE) received the master's degree from Lund University, Sweden, and the Ph.D. degree from KTH. He is a Professor of wireless communication with KTH Royal Institute of Technology, Sweden, where he also leads the Department of Communication Systems and SweWIN Research Center. He has coauthored four textbooks, published 400 peer-reviewed papers, created 70 open-source software repositories, and is a Clarivate Highly Cited Researcher.\par\medskip

\noindent\textbf{Sunwoo Kim} (Senior Member, IEEE) received the B.S. degree from Hanyang University, Seoul, Korea, in 1999, and the Ph.D. degree from the Department of Electrical and Computer Engineering, University of California, Santa Barbara, in 2005. Since 2005, he has been with the Department of Electronic Engineering, Hanyang University, Seoul, Korea, where he is currently a Professor. He is the Director of the Beyond-G Global Innovation Center, and a former Director of the 5G/Unmanned Vehicle Research Center, funded by the Ministry of Science and ICT of Korea. He is also a Vice President of the Korean Institute of Communications and Information Sciences (KICS) and has been a visiting scholar at the Laboratory for Information and Decision Systems, Massachusetts Institute of Technology, in 2018--2019 and 2023. He served as an Associate Editor of IEEE Transactions on Vehicular Technology. His research interests include wireless communication, sensing, and localization.\par\medskip

\noindent\textbf{Marios Kountouris} (Fellow, IEEE) received the diploma degree in electrical and computer engineering from the National Technical University of Athens (NTUA), Greece, and the M.S. and Ph.D. degrees in electrical engineering from T\'{e}l\'{e}com ParisTech, France. He is a Professor at \mbox{EURECOM}, France, and a Distinguished Researcher at the University of Granada, Spain. He has published over 230 peer-reviewed papers and is a co-inventor of more than 10 patents.\par\medskip

\end{document}